\documentstyle[preprint,aps,epsf]{revtex}
\begin{document}
\begin{titlepage}
\title{On the equivalence of
regularization schemes \footnote{Work supported in part by the
National Nature Science Foundation of China under Grant No.
10075020.}}
\author{Ji-Feng Yang}
\address{Department of Physics, East China Normal University,
Shanghai 200062, P R China} \maketitle
\begin{abstract}
We illustrated via the sunset diagram that dimensional
regularization 'deforms' the nonlocal contents of multi-loop
diagrams with its equivalence to cutoff regularization scheme
recovered only after sub-divergence were subtracted. Then we
employed a differential equation approach for calculating loop
diagrams to verify that dimensional regularization deformed the
'low energy' contents before subtraction. The virtues of the
differential equation approach were argued especially in
nonperturbative perspective.
\end{abstract}
\end{titlepage}
\section{Introduction}
Recently, there have been intensive applications of the effective
field theory method\cite{EFT} to nucleon interactions\cite{Nucl}.
In these applications UV divergences occur and are subtracted
following the field theoretical renormalization principles.
However, since the employed framework is in fact nonperturbative
ones, there appear some subtle issues like the applicabilities of
various regularization and/ or subtraction schemes\cite{QMDR}. It
is recently shown that the dimensional regularization is
equivalent to cutoff scheme in the only after it is combined with
some nonminimal subtraction scheme\cite{Scheming}.

In this report, we discuss the subtleties associated with the
conventional regularization schemes. We wish to emphasize that
dimensional regularization is inequivalent to cutoff
regularization in parametrizing the nonlocal contents of
multi-loop diagrams \emph{before} any subtraction is done, and
that the equivalence is only recovered with a sub-divergence
subtraction that affected the definite nonlocal part, in contrast
to the cutoff scheme situation. On the other hand the cutoff
scheme is notorious for its power divergences. Although these
facts are already known to high energy theorists, we still feel it
necessary to point out that the renormalization is easily operable
within the perturbative diagrammatic contexts, but not necessarily
so in various (known and unknown) nonperturbative contexts. Thus
we should be aware of this subtlety when working in
nonperturbative contexts.

This report is organized as follows. In section two, we exhibit
the obviously different results calculated respectively through
dimensional and cutoff regularization. The difference in the
leading definite part is removed after the sub-divergence is
removed. Then, in section three we apply a differential equation
approach to calculate the two loop integral and find that it is
compatible with the cutoff result but not with the dimensional
regularization result before subtraction. The last section is
devoted to discussion and the summary.

\section{The sunset diagram in massive $\lambda\phi^4$}
First let us specify the lagrangian 
\begin{equation}
L\equiv\frac{1}{2}(\partial\phi)^2-\frac{1}{2}m^2\phi^2-
\frac{\lambda}{4!}\phi^4.
\end{equation}
To focus on the mass dependence we put the external momentum to
zero. The two loop self-energy diagram under consideration is
\begin{eqnarray}
{\Sigma}^{(2)}_{\theta}
(0,m^2)\equiv-\frac{\lambda^2}{6}I_{\theta}(m^2)\equiv
-\frac{\lambda^2}{6}\int\int \frac{d^4k
d^4l}{(2\pi)^8}\frac{1}{(k^2+m^2)((k+l)^2+m^2)(l^2+m^2)}
\end{eqnarray}
where we have Wick-rotated all the internal momentum integrals.
This is a nontrivial two loop diagram with overlapping
divergences. This diagram has been calculated a number of times in
various regularization schemes\cite{Jackiw,Collins,FJ}, we can
extract from these papers the expressions for $I_{\theta}(m^2)$.
First let us see the dimensional regularization result.
\subsection{Dimensional regularization}
From Ref.\cite{Collins} we find that in $D$ dimensions
\begin{eqnarray}
I_{\theta}^{(D)}(m^2)=\frac{\Gamma(3-D)}{(4\pi)^{2D}}\int^1_0
[d\alpha d\beta d\gamma]
\delta(\alpha+\beta+\gamma-1)\frac{(m^2(\alpha+\beta+\gamma))^{D-3}}
{(\alpha\beta+\beta\gamma+\gamma\alpha)^{D/2}},\end{eqnarray} and
from \cite{FJ} we have
\begin{eqnarray}
&&I_{\theta}^{(D)}(m^2)=
-\frac{2\Gamma(2-D/2)\Gamma(1-D/2)m^{2D-6}}{(4\pi)^D}X(1),\nonumber
\\
&&X(1)\equiv \frac{2}{3(3-D)}+3^{(D-3)/2} \int^1_0 du
 \frac{u^{2-D/2}/(4-u)-1}{u^{1/2}(4-u)^{(D-1)/2}}.
\end{eqnarray}
The full $\epsilon$-expansion is given in \cite{FJ} as
\begin{eqnarray}
(4\pi)^4 I_{\theta}^{(D)}(m^2)=-\frac{3m^2}{2\epsilon^2}
(1+(3-2\overline{L}(m))\epsilon+a\epsilon^2+\cdots)
\end{eqnarray}
with $\overline{L}(x)\equiv\ln(x^2/\mu^2)+\gamma_E-\ln4\pi$,
$\gamma_E$ Euler's constant and
\begin{eqnarray}
a=2\overline{L}^2(m)-6\overline{L}(m))+7
+\zeta(2)+\frac{1}{\sqrt{3}}\int^{\pi/6}_0 dx \ln(2\sin x).
\end{eqnarray}
This result is the same as Collins' after carrying out the
integrals in Eq.(3)
\begin{eqnarray}
(4\pi)^4 I_{\theta}^{(D)}(m^2)=-\frac{3m^2}{2}\{
\frac{1}{\epsilon^2}+\frac{3-2\overline{L}(m)}{\epsilon}
-6\overline{L}(m)+2\overline{L}^2(m)+const.\}.
\end{eqnarray}
Now we turn to the cutoff scheme.
\subsection{Cutoff regularization}
This two loop integral calculated in cutoff scheme can be
extracted from Ref.\cite{Jackiw}
\begin{eqnarray}
(4\pi)^4I_{\theta}^{(\Lambda)}(m^2)=-\frac{3m^2}{2}\{\ln^2
(m^2/\Lambda^2)+2\ln(m^2/\Lambda^2)\}+C \Lambda^2
\end{eqnarray} with the coefficient $C$ of $\Lambda^2$ term
not explicitly given there. Through direct calculation $C$ can be
determined as $ C=2$.

In order to make the comparison easier, we cast the cutoff result
into the following form
\begin{eqnarray}
(4\pi)^4I_{\theta}^{(\Lambda)}(m^2)= \frac{-3m^2}{2} \{
\overline{L}^2(\Lambda)+\overline{L}(\Lambda)(2-2\overline{L}(m))
-2\overline{L}(m)+\overline{L}^2(m) \} +2\Lambda^2.
\end{eqnarray}
\subsection{Disadvantages of conventional regularization schemes}
Now it is clear that the leading finite term (a double log
$\overline{L}^2(m)$) that has \emph{different coefficients in
different schemes}, i.e.,
$-\frac{3m^2}{(4\pi)^4}\overline{L}^2(m)$ in dimensional
regularization and  $-\frac{3m^2}{2(4\pi)^4}\overline{L}^2(m) $ in
cutoff scheme. This double log is the leading definite part of the
sunset diagram that should be independent of the UV
regularizations residing in the local part. Thus subtraction of
divergence (including sub-divergence) should not alter the leading
finite terms(nonlocal), or the leading finite term is
renormalization/subtraction scheme independent. But in dimensional
scheme, the sub-divergence subtraction does affect this double log
as can be seen from the counter-term in Ref.\cite{FJ}:
\begin{eqnarray}
&&I^{(D)\prime}_{\theta}(m^2)\equiv I^{(D)}_{\theta}(m^2)-
\frac{3}{(4\pi)^2\epsilon}I^{(D)}_1(m^2)\nonumber \\
&=&I^{(D)}_{\theta}(m^2) +\frac{3}{(4\pi)^4 \epsilon^2}
\{1+(1-\overline{L}(m))\epsilon+(\frac{\overline{L}^2(m)}{2}
-\overline{L}(m)+1+\frac{\zeta(2)}{2})\epsilon^2+\cdots\}\nonumber
\\ &=&-\frac{3m^2}{2(4\pi)^4\epsilon^2}\{-1+\epsilon+(\overline{L}^2(m)-
4\overline{L}(m))\epsilon^2+\cdots\}
\end{eqnarray} where $I^{(D)}_1 (m^2)\equiv\int\frac{\mu^{2\epsilon}d^D
k}{(2\pi)^D}\frac{1}{k^2+\Omega^2}$ denotes the tadpole
sub-diagram. It is obvious that the leading double log now is in
agreement with cutoff scheme result as the cutoff scheme
sub-divergence subtraction will not affect the double log at all.
It is also obvious that the counter-term \emph{did} subtract the
leading nonlocal term in dimensional regularization.

The origin of the problem lies in the deformation of the low
energy (nonlocal) content of a diagram in the dimensional
regularization that \emph{should be independent of regularization
schemes} as it is purely 'low energy' part that should be
unaffected by the 'high energy' details. Although we can remove it
through sub-diagram subtractions within perturbative context, we
feel that it is a subtle and disadvantageous aspect of dimensional
regularization that might cause problems in other more complicated
contexts than the perturbative one.

One might argue in favor of dimensional regularization that there
is no notorious quadratic and quartic divergences in it. In our
opinion, the vanishing of power form divergences in dimensional
regularization might be understood as an implicitly built-in
subtraction, while such subtraction must be explicitly done in
cutoff scheme. It is more natural to prefer the ability to produce
the correct nonlocal (or low energy) information. However, more
explicit divergences did make their removal more difficult in
nonperturbative contexts. It is demanding to find a scheme to get
rid of these weak points. This is what we wish to do in next
section. In addition, we mention that in dimensional scheme the
definition of metric tensor and the Dirac algebra becomes a rather
nontrivial and complicated task, especially in presence of
fermions.
\section{A differential equation analysis of loop diagrams}
Now let us calculate the diagram by adopting the standard point of
view that all the known QFT's are effective theories for a
completely well-defined quantum theory containing all the
underlying high energy details\cite{PS2}. Since we have not found
the complete theory yet, usually we introduce some artificial
deformations to make QFT's UV finite that might be physically
inappropriate, for example dimensional regularization deforms the
low energy parts that must be recovered through subtraction in the
perturbative context.

However, it is well known fact that differentiation with respect
to physical parameters (usually external momenta and/or masses)
reduces the divergence degrees (or more rigorously the
ill-definedness degrees) of the divergent diagrams\cite{CK}. We
can perform the differentiation on a diagram $\gamma$ for enough
times (for $\omega_{\gamma}+1$ times with the divergence degree
$\omega_{\gamma}$) to yield a sum of diagrams without any
superficial ill-definedness. If there is still sub-divergence,
repeat the operation on the sub-diagrams till no ill-definedness
is left so that the loop integrals can be safely carried out.
Finally we integrate back with respect to the momenta (and/or
masses) 'external' to the loop treated, which can be done diagram
by diagram and loop by loop\cite{CGR}.

Technically, this approach amounts to calculating the ill-defined
loop diagrams by solving of well defined differential equations
derived from the \emph{existence} of the underlying theory and its
'low energy' limits, a generalization of the WT identity in gauge
theories. \emph{The ill-defined diagrams or loops should be well
defined in the underlying theory} with the ill-definedness
indicating the lacking of necessary UV details in the effective
theories or the incompleteness of the effective theories. Then in
this approach the solutions would naturally contain unknown
constants parametrizing the ill-definedness or incompleteness (to
be fixed by physical 'boundary conditions') in contrast to the
cutoff and dimensional schemes where \emph{ill-definedness is
parametrized in terms of divergences}, which is an eminent
progress in regularization. It is obvious that this approach needs
neither artificial deformations nor unnatural cutoffs. And the
metric tensor and Dirac algebra need no more additional treatment
here.

Now we apply this method to the diagram discussed in last section,
that is to determine the two loop integral $I_{\theta}(m^2)$.

(1). First, we differentiate it with respect to mass square
($m^2$) for two times (differentiation with respect to momentum
could achieve the same goal), that is,
\begin{eqnarray}
&&\frac{\partial^2}{\partial_{(m^2)^2}}I_{\theta}(m^2)\equiv
6I_{\theta:(3;1;1)}(m^2)+3I_{\theta:(2;2;1)}(m^2)+3I_{\theta:(2;1;2)}(m^2),\\
&&I_{\theta:(\alpha;\beta;\gamma)}(m^2)\equiv \int \frac{d^4k
d^4l}{(2\pi)^8(k^2+m^2)^{\alpha}((k+l)^2+m^2)^{\beta}(l^2+m^2)^{\gamma}}.
\end{eqnarray}
The result is a sum of three new diagrams without any overall
divergence. Among these diagrams, $I_{\theta:(2;2;1)}(m^2)$ and
$I_{\theta:(2;1;2)}(m^2)$ contain no sub-divergence while $
I_{\theta:(3;1;1)}(m^2)$ contains a sub-divergence in the
integration of loop momentum $l$
\begin{eqnarray}
I_{(1;1)}(m^2,k^2)\equiv \int
\frac{d^4l}{(2\pi)^4((k+l)^2+m^2)(l^2+m^2)}.
\end{eqnarray}

(2). Second, we treat this divergent sub-diagram with the same
method described above which lead to the following inhomogeneous
differential equation
\begin{eqnarray}
\partial_{m^2}I_{(1;1)}(m^2,k^2)&=&-\int
\frac{d^4l}{(2\pi)^4}\frac{((k+l)^2+m^2)+(l^2+m^2)}
{((k+l)^2+m^2)^2(l^2+m^2)^2}\nonumber \\
&=&\frac{-1}{(4\pi)^2}\int^1_0 \frac{dx}{m^2+(x-x^2)k^2}.
\end{eqnarray}
The solution to this equation is easy to find,
\begin{eqnarray}
I_{(1;1)}(m^2,k^2)=\frac{-1}{(4\pi)^2}\int^1_0 dx
\{\ln(m^2+(x-x^2)k^2)+c_{(1:1)}\}
\end{eqnarray}with $c_{(1;1)}$ being the integration constants
independent of any physical parameters like masses and momenta. In
specific regularization schemes, $c_{(1;1)}$ are taken place by
various constants containing divergent ones, e.g., $\ln\Lambda^2$
in cutoff scheme. In principle, it should be fixed by 'boundary
condition'.

(3). Now we can compute the right hand side of Eq.(11) and obtain
again an inhomogeneous differential equation as below
\begin{eqnarray}
\frac{\partial^2}{\partial_{(m^2)^2}}I_{\theta}(m^2)=
-\frac{3}{(4\pi)^4} \frac{\ln m^2+c_{(1;1)}-1}{m^2},
\end{eqnarray} and the solution to it reads
\begin{eqnarray}
I_{\theta}(m^2)=-\frac{3m^2}{2(4\pi)^4}\{\ln^2 m^2
+2(c_{(1;1)}-2)\ln
m^2-2c_{(1;1)}+4+2c_{\theta;1}\}-\frac{3c_{\theta;2}}{(4\pi)^4}
\end{eqnarray}with $c_{\theta;1},c_{\theta;2}$ being the
constants (independent of masses, coupling and momenta) to be
fixed by 'boundary conditions', as pointed out in the preceding
paragraghs.

It is clear that the leading double log
($-\frac{3m^2}{2(4\pi)^4}\ln^2 m^2$) determined from the
differential equation Eq.(16) naturally satisfied by
$I_{\theta}(m^2)$ agrees with that determined from cutoff scheme
but disagrees with that from dimensional regularization. In fact
we can define the constants $c_{(1;1)}, c_{\theta;1}$ and
$c_{\theta;2}$ in such a way that the cutoff result is exactly
reproduced, i.e., the cutoff scheme can be a particular solution
to the differential equations while dimensional regularization is
not. Thus, dimensional regularization is disfavored in the sense
just specified. We stress again that this is a property
\emph{before} subtraction.

As was already mentioned above, dimensional regularization
'deforms' the low energy information of a quantum theory that must
be recovered through a necessary subtraction of some low energy
contents, while the cutoff scheme preserves the low energy
information faithfully but yields some ugly divergences. In this
connection, our differential equation approach is more favorable.
It should show particular advantage in nonperturbative contexts
since the troublesome work of subtracting divergences
(perturbative and/or nonperturbative) is replaced by the easier
work of fixing the undetermined constants through physical
'boundary conditions'. Moreover, we need not the extra recourse to
the tricky procedures for manipulating infinities and the
intermediate renormalized 'quantities' that must be finally
translated into physical 'quantities'. It is not difficult to see
that in certain nonperturbative problems, to parameterize the
ill-definedness in terms of nonperturbative divergences might make
the subtraction procedure extremely complicated and awkward might
even lead to no useful (or trustworthy) predictions which is
physically unfavorable, while to parametrize the ill-definedness
in terms of unknown constants will incur no awkwardness and make
physical predictions more available\cite{NILOU}, a superiority
over the conventional schemes.
\section{Discussions and Summary}
Where is the technical source of the subtlety for dimensional
regularization? We feel that it is in the special way that
dimensional regularization parameterizes the finite nonlocal
piece, it is in the functional form of $(m^2 or p^2)^{-n\epsilon}$
and its expansion in terms of $\epsilon$ in the $n$-loop
integrals. It is mathematically sound to expand
$(m^2)^{-2\epsilon}$ as
\begin{eqnarray}
(m^2)^{-2\epsilon}=\exp(-2\epsilon\ln m^2)=
\sum^{\infty}_{n=0}\frac{(-2\epsilon)^n}{n!}=1-2\epsilon\ln
m^2+2\epsilon^2 \ln^2 m^2+o(\epsilon^3).
\end{eqnarray}But the $\epsilon^2$ order term has a 'incorrect'
numeric coefficient 2. The double log should vanish in the tadpole
diagram when the latter is not used to subtract the
sub-divergences in the two-loop diagram, but should be kept when
used for removing sub-divergence.

If one took the $(m^2)^{-2\epsilon}$ as a product of two
$(m^2)^{-\epsilon}$ with all higher order ($\epsilon^n, n\geq2$)
terms dropped, one would get the correct double log, i.e.,
\begin{eqnarray}
(m^2)^{-2\epsilon} \equiv \{(m^2)^{-\epsilon} \}^2 \equiv \{
1-\epsilon \ln m^2 \}^2=1-2\epsilon \ln m^2 + \epsilon^2 \ln^2
m^2.
\end{eqnarray}
Unfortunately, this 'one-loop' convention has at least two
problems: (a). One has to specify how to expand other factors that
are functions of $\epsilon$ in a manner that is consistent with
this 'one-loop' convention, and one can find that there can be
several ways to expand the factors that will yield different
numerical constants; (b). The more serious problem is that when
one makes dimensional regularization and cutoff scheme agree with
each other on the double log in this two loop diagram, they will
disagree with each other on other parts and on the other diagrams
(and the subtraction of sub-divergence must be carefully defined).

Although the problem is analyzed in a perturbative example, our
main concern is in the difficulties one might meet in
nonperturbative contexts. Since nonperturbative investigations in
standard model and symmetry breaking physics are inevitable, we
feel it necessary to be aware of such regularization subtleties
and to work out a more efficient scheme. We hope our discussions
here might of some value both in the pure theoretical perspective
and in the application perspective.

In summary, the subtleties in the regularization scheme
equivalence was discussed through the sunset diagram in massive
$\lambda\phi^4$ theory. We introduced a natural differential
equation approach for calculating the loop diagrams to help the
analysis and demonstrated that the widely used dimensional
regularization deformed the 'low energy' contents of theory before
subtraction is made in contrast to the cutoff scheme. Our
differential equation approach showed more virtues comparing to
the two widely used regularization schemes.

\section*{Acknowledgement}
The author is grateful to W. Zhu for helpful conversations.

\end{document}